\documentclass[aps,prl,twocolumn,superscriptaddress]{revtex4}
\usepackage{graphicx}
\usepackage{dcolumn}
\usepackage{bm}
\usepackage{verbatim}
\usepackage{color}
\usepackage{soul}
\usepackage[colorlinks]{hyperref}
\input{epsf}

\usepackage[sort&compress]{natbib}
\begin{document}


\title{Effect of rotation of the polarization of linearly polarized microwaves on the radiation-induced
magnetoresistance oscillations}
\author{A. N. Ramanayaka}

\author{R. G. Mani}
\affiliation{Department of Physics and Astronomy, Georgia State
University, Atlanta, GA 30303}

\author{J. I\~narrea}
\affiliation {Escuela Polit\'ecnica Superior,Universidad Carlos
III,Leganes,Madrid,28911,Spain}

\author{W. Wegscheider}
\affiliation{Laboratorium f\"{u}r Festk\"{o}rperphysik, ETH
Z\"{u}rich, 8093 Z\"{u}rich, Switzerland}

\date{\today}
\begin{abstract}
Light-matter coupling is investigated by rotating, by an angle
$\theta$, the polarization of linearly polarized microwaves with
respect to the long-axis of GaAs/AlGaAs Hall-bar electron devices.
At low microwave power, $P$, experiments show a strong sinusoidal
variation in the diagonal resistance $R_{xx}$ vs. $\theta$ at the
oscillatory extrema, indicating a linear polarization sensitivity
in the microwave radiation-induced magnetoresistance oscillations.
Surprisingly, the phase shift $\theta_{0}$ for maximal oscillatory
$R_{xx}$ response under photoexcitation appears dependent upon the
radiation-frequency $f$, the extremum in question, and the
magnetic field orientation or $sgn(B)$.
\end{abstract}
\maketitle

\section{introduction}

Vanishing electrical resistance  has long been viewed as a
harbinger of new physics in condensed matter since the discovery
of superconductivity.\cite{grid-5} Transport studies of
two-dimensional electron systems (2DES) supported this notion by
revealing the quantum Hall effects as correlates of vanishing
diagonal resistance  at low temperatures, $T$, and high magnetic
fields, $B$.\cite{grid-2} In the recent past, low-$B$ transport
studies under microwave irradiation in the 2DES uncovered the
possibility of eliminating backscattering by photo-excitation,
without concurrent Hall quantization.\cite{grid1,grid2} The
experimental realization of such radiation-induced zero-resistance
states, and associated $B^{-1}$-periodic radiation-induced
magnetoresistance oscillations expanded the experimental\cite{
grid1, grid2, grid2a, grid2b, grid2c, grid107, grid5, grid6,
grid7, grid9, grid12, grid13, grid20, grid22, grid22b, grid22d,
grid22c, grid22e} and theoretical\cite{grid23, grid24, grid25,
grid101, grid27, grid33, grid34, grid37, grid102, grid38, grid43,
grid46, grid48, grid49, grid103, grid104} investigations of
light-matter coupling in low-dimensional electronic systems.
Indeed, microwave excitation of semiconductor quantum wells and
graphene ribbons is now viewed as an approach to artificially
realizing a (Floquet) topological insulator for possible
applications in topological quantum computing and
spintronics.\cite{grid103, grid104, grid105, grid106}

Microwave-induced zero-resistance states appear when the
associated $B^{-1}$-periodic magnetoresistance oscillations grow
in amplitude and become comparable to the dark resistance of the
2DES. Such oscillations, which exhibit nodes at cyclotron
resonance and harmonics thereof,\cite{grid1, grid2a} are now
understood via the the displacement model,\cite{grid23, grid25,
grid27, grid46} the non-parabolicity model,\cite{grid101} the
inelastic model,\cite{grid33} and the radiation driven electron
orbit model.\cite{grid34, grid37, grid102} Recently, a
magneto-plasmon approach has also been motivated.\cite{grid49} In
theory, some of these mechanisms can drive the magnetoresistivity
to negative values at the oscillatory minima. Negative resistivity
then triggers an instability in favor of current domain formation,
and zero-resistance states.\cite{grid24, grid43}

A distinguishing feature between existing theories for the
radiation-induced oscillating magnetoresistivity concerns the role
of the microwave-polarization. Here, the displacement model
predicts that the oscillation-amplitude depends on whether the
linearly polarized microwave electric field, $E_{\omega}$, is
parallel or perpendicular to the $dc$-electric field,
$E_{DC}$.\cite{grid25} More specifically,\cite{grid25}, the
inter-Landau level contribution to the photo-current in this
theory includes a term with a Bessel function whose argument
depends upon whether $E_{DC}$ and $E_{\omega}$ are parallel or
perpendicular to each other, and this Bessel-function-argument is
a constant for circular polarized or unpolarized radiation for any
ratio of $\omega_c/\omega$.\cite{grid25} In contrast, the
inelastic model suggests insensitivity of the photoconductivity to
the polarization orientation of the linearly polarized microwave
field.\cite{grid33} The radiation-driven electron orbit model
indicates a polarization immunity that depends upon the damping
factor, $\gamma$,-  a material- and sample-dependent parameter,-
exceeding the microwave angular frequency, $\omega$.\cite{grid37,
grid102} Finally, within the non-parabolicity model, the effect of
irradiation on $dc$ transport emerges only for linear-, but not
circular,- polarization of the radiation field.\cite{grid101}
Consequently, the radiation induced contribution within this
theory depends on the relative orientation between $E_{DC}$ and
the linearly polarized $E_{\omega}$.\cite{grid101}

The polarization aspect has been explored by experiment in
ref.\cite{grid5}, \cite{grid13}, and \cite{grid22e}. Measurements
carried out on L-shaped specimens have suggested that the period
and phase of the radiation induced magnetoresistance oscillations
are the same for the $E_{\omega} \parallel I$ and $E_{\omega}
\perp I$ configurations.\cite{grid5} Ref.\cite{grid13} has
reported the immunity of microwave-induced magneto-resistance
oscillations and zero resistance states to the sense of circular-
and linear- polarizations from the experiments carried out on
specimens with a square geometry in a quasioptical setup. In a
recent study, Mani et al. \cite{grid22e} reported a strong
sensitivity in the amplitude of the radiation-induced
magnetoresistance oscillations to the relative orientation of the
linear polarization with respect to the Hall bar axis.

Here, we examine the effect of rotating the polarization of
linearly polarized microwaves on the radiation-induced
magnetoresistance oscillations in the GaAs/AlGaAs 2D electron
system.\cite{grid22e} Surprisingly, at low microwave power, $P$,
experiments indicate a strong sinusoidal response as $R_{xx}
(\theta) = A \pm C \cos^{2}(\theta - \theta_{0})$ vs. the
polarization rotation angle, $\theta$, with the $'+'$ and $'-'$
cases describing the maxima and minima, respectively. At higher
$P$, the principal resistance minimum exhibits additional extrema
vs. $\theta$. Notably, the phase shift $\theta_{0}$ can vary with
$f$, $B$, and $sgn(B)$.

\section{Experiment and Results}

These polarization-dependence studies utilized the novel setup
illustrated in Fig. 1(a). Here, a rotatable MW-antenna introduces
microwaves into a $11$ mm. diameter circular waveguide. The
circular symmetry then allows the rotation of the antenna and the
polarization with respect to the stationary sample, see Fig. 1(a)
and 1(b). Note that the transverse electric (TE) mode is excited
by the microwave (MW) antenna of Fig. 1(a), and the specimen is
subject to the $TE_{11}$ mode of the circular waveguide as shown
in Fig. 1(c). These scaled sketches of the small (0.4 mm wide)
Hall bar sample within the 11 mm i.d. waveguide, with superimposed
electric field lines (see Fig. 1(c)), suggest a well defined
polarization over the active area of the specimen, for all
rotation angles. The samples consisted of $400\mu m$-wide Hall
bars characterized by n (4.2K) = 2.2$\times$$10^{11}$ $cm^{-2}$
and $\mu \approx 8\times 10^{6}cm^{2}/Vs$. The long axis of these
Hall bars were visually aligned parallel to the polarization axis
of the MW-antenna for the data exhibited in Fig. 1 - 4, and this
defined $\theta=0^{0}$. Thus, $\theta$, see Fig. 1(b) and Fig.
1(c) represents the polarization rotation angle. Note that, for
the measurements exhibited in Fig. 5, however, the Hall bar was
oriented perpendicular to the MW-antenna at the outset, i.e., at
$\theta = -90^{0}$.

\begin{figure}[t]
\begin{center}
\leavevmode \epsfxsize=2.1 in \epsfbox {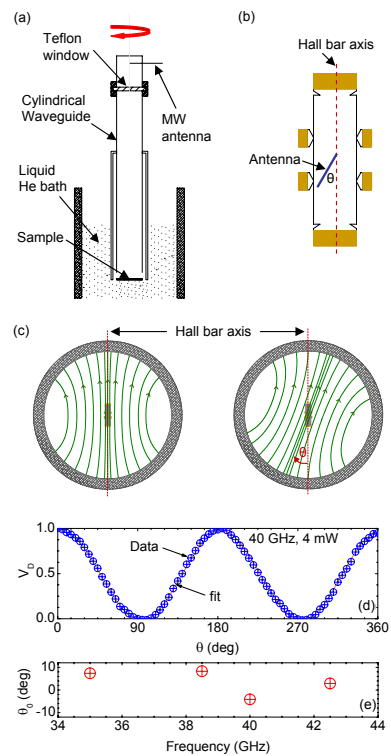}
\end{center}
\caption{\label{fig:epsart} (color online) (a) A microwave (MW)
antenna is free to rotate about the axis of a cylindrical
waveguide.  (b) A Hall bar specimen, shown as ``sample" in (a), is
oriented so that the Hall bar long-axis is parallel to the
MW-antenna for $\theta = 0^{0}$. (c) This scaled figure shows the
$TE_{11}$ mode electric field pattern within the waveguide with
the Hall bar superimposed on it. The left panel illustrates
$\theta = 0^{0}$ case, while the right panel shows the finite
$\theta$ case. Note the parallel electric field lines within the
active area of the specimen. (d) shows the normalized response
($V_{D}$) of the diode detector (circles) placed at the sample
position,  at $f = 40 GHz$. (e) The phase shift, $\theta_{0}$,
obtained from a fit to $V_{D}$ vs. $\theta$ is shown vs. the
radiation frequency, $f$.}\end{figure}

\begin{figure}[t]
\begin{center}
\leavevmode \epsfxsize=2.1 in \epsfbox {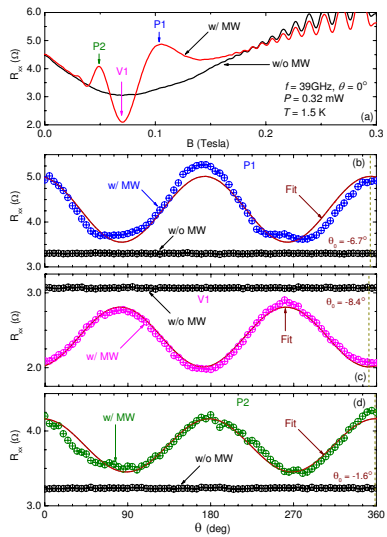}
\end{center}
\caption{\label{fig:epsart} (color online) (a) The dark- and
microwave excited- magnetoresistance $R_{xx}$ are exhibited. Here,
the microwave antenna is parallel to the long axis of the Hall
bar, i.e., $\theta = 0$. The principal maxima have been labelled
$P1$ and $P2$, and the minimum is $V1$. (b), (c), and (d) show the
experimental extremal $R_{xx}$ response at $P1$, $V1$, and $P2$,
respectively. (b) and (d) show that, at the maxima $P1$ and $P2$,
$R_{xx}$ under photoexcitation exceeds the dark $R_{xx}$. On the
other hand, at $V1$, the $R_{xx}$ under photoexcitation lies below
the dark $R_{xx}$. }
\end{figure}

\begin{figure}[t]
\begin{center}
\leavevmode \epsfxsize=2.5 in \epsfbox {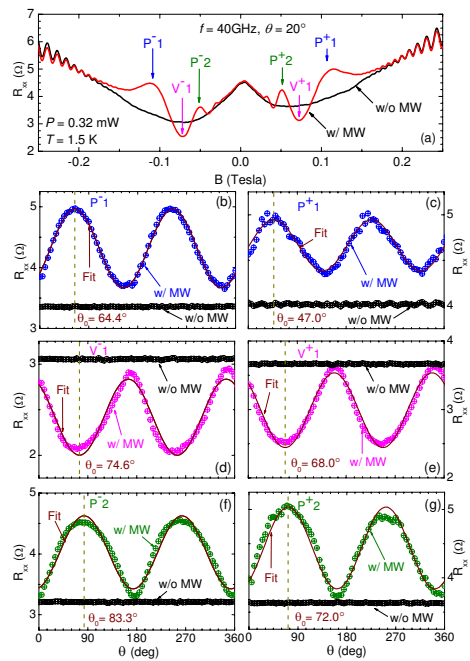}
\end{center}
\caption{\label{fig:epsart} (color online) This figure compares
the angular response for positive and negative magnetic fields.
(a) Dark- and photo-excited- $R_{xx}$ are shown at $f = 40 GHz$
with $\theta = 20^{0}$ over the $B$-range $-0.25 \le B \le 0.25T$.
(b), (d), and (f) show the $\theta$ dependence of $R_{xx}$ of the
principal maxima $P^{-} 1$ (b), $P^{-} 2$ (f), and the minimum
$V^{-} 1$ (d) for $B < 0$. (c), (e), and (g) show the $\theta$
dependence of $R_{xx}$ of the principal maxima $P^{+} 1$ (c),
$P^{+} 2$ (g), and the minimum $V^{+} 1$ (e) for $B > 0$. In these
figures, the w/o MW traces indicate the sample response in the
dark, while the w/ MW traces indicate the response under
photo-excitation. The phase shift, $\theta_{0}$, is indicated by a
vertical dashed line in (b) - (g).}
\end{figure}

\begin{figure}[t]
\begin{center}
\leavevmode \epsfxsize=2.1 in \epsfbox {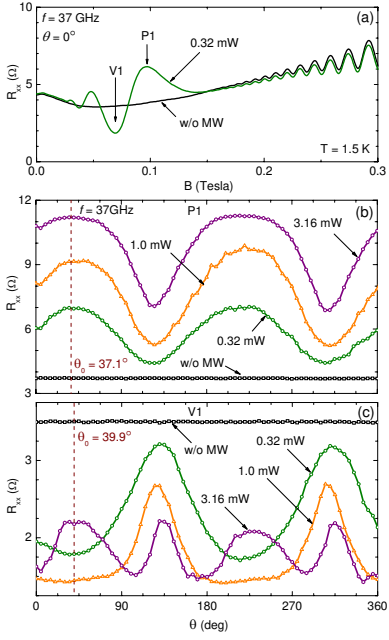}
\end{center}
\caption{\label{fig:epsart} (color online) This figure examines
the angular response of $R_{xx}$ at different microwave power
levels, $P$. (a) Magnetoresistance oscillations in $R_{xx}$ are
exhibited for $f = 37 GHz$ with $\theta = 0$ and $P = 0.32 mW$,
along with the dark $R_{xx}$ curve. (b) shows the $\theta$
dependence of $R_{xx}$ of the principal maximum $P1$. (c) shows
the $\theta$ dependence of $R_{xx}$ of the principal minimum $V1$.
In these figures, the w/o MW (w/ MW) traces indicate the sample
response in the absence (presence) of microwave photo-excitation.
Note that, in (c), additional peaks occur near $\theta = 45^{0}$
and $\theta = 225^{0}$ at $P=3.16 mW$.}
\end{figure}

\begin{figure}[t]
\begin{center}
\leavevmode \epsfxsize=2.1 in \epsfbox {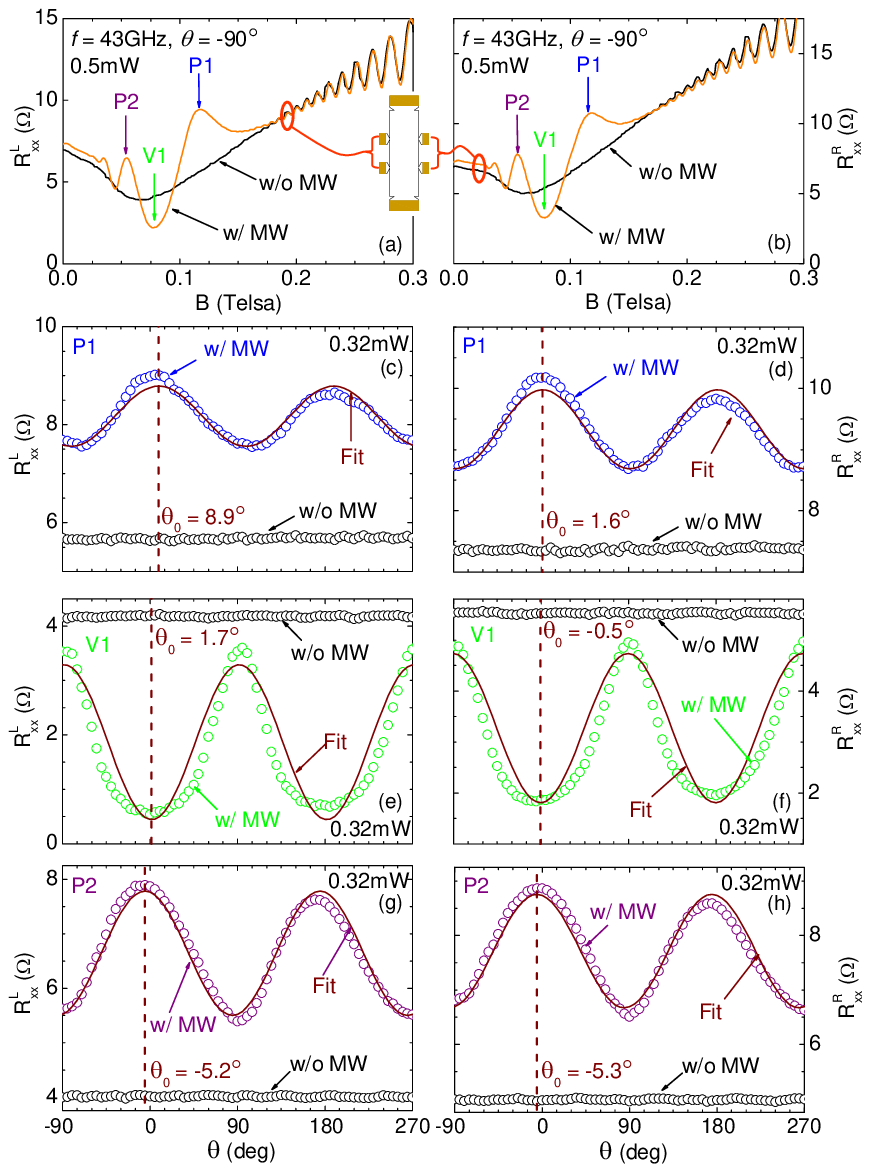}
\end{center}
\caption{\label{fig:epsart} (color online) This figure exhibits
the angular dependence of the diagonal resistance on the left and
right sides of the Hall bar device, see panel (a) inset.  Dark-
and photo-excited- $R_{xx}$ are shown for the (a) left side,
$R^{L} _{xx}$, and (b) right side, $R^{R} _{xx}$, at $f=43GHz$
with $P=0.5mW$ and $\theta = -90^{0}$. Panels (c), (e), and (g)
show the $\theta$ dependence of the $R^{L} _{xx}$ for $P=0.32mW$
at the first maximum (P1), first minimum (V1), and second maximum
(P2), respectively. Similarly, panels (d), (f), and (h) show the
$\theta$ dependence of the $R^{R} _{xx}$ for $P=0.32mW$ at P1, V1,
and P2, respectively. Phase shifts obtained for the two sides of
the Hall bar, $R^{L} _{xx}$ and $R^{R} _{xx}$, show similar values
within the experimental uncertainty at $f=43GHz$.}
\end{figure}


Tests of this setup utilized also an ``analyzer" consisting of a
probe-coupled antenna and a square law detector. Measurements
carried out with the MW-antenna (Fig. 1(a)) connected directly to
the analyzer indicated that polarized microwaves were generated by
the MW-antenna. In the next step, the waveguide sample holder was
inserted between the polarizer (MW-antenna) and the analyzer.
Here, the analyzer was fixed at a particular orientation, and the
MW-antenna was rotated over $360^{0}$. Fig. 1(d) shows the
normalized detector response, $V_{D}$, at $f = 40 GHz$. Figure
1(d) exhibits the expected sinusoidal variation, i.e., $V_{D}
\propto \cos^{2}{\theta}$, for linearly polarized radiation, as a
function $\theta$. Also shown in Fig. 1(d) is a fit to $V_{D} = A
+ C\cos^2(\theta - \theta_{0})$ that is used to extract
$\theta_{0}$. Fig. 1(e) shows the variation of $\theta_{0}$ with
the frequency, $f$, with the analyzer in place of the specimen.
Here, $\theta_{0} \le 10^{0}$ for $34 \le f \le 44 GHz$. This
result shows that, without the sample, the polarization at the
sample-location follows expectations, within an experimental
uncertainty of approximately $10^{0}$.

Figure 2 (a) shows the dark- and photo-excited- diagonal
resistance $R_{xx}$ vs. $B$. Here, the photo-excited measurement
was carried out with microwave frequency $f = 39 GHz$ and
microwave power $P = 0.32 mW$, and the MW-antenna parallel to the
Hall bar long-axis, i.e., $\theta = 0^{0}$. Fig 2(a) shows once
again a well-known negative magnetoresistance to $B  = 0.075$
Tesla in the high mobility specimen in the dark
condition.\cite{grid107} In Fig. 2(a), the labels $P1$, $V1$, and
$P2$, identify the oscillatory extrema that are examined in Fig.
2(b), (c), and (d), respectively. Fig. 2(b) and (d) show that the
photo-excited $R_{xx}$ (i.e, ``w/ MW") traces lie above the dark
(i.e., ``w/o MW") $R_{xx}$, traces at the resistance maxima for
all $\theta$. Further, the photo-excited $R_{xx}$ at $P1$ and $P2$
fits the function $R_{xx} (\theta) = A + C \cos^{2} (\theta -
\theta_{0})$, with $\theta_{0} = -6.7^{0}$ and $-1.6^{0}$,
respectively. Fig. 2(c) shows that, at the resistance minimum
$V1$, the w/ MW $R_{xx}$ trace lies below the dark $R_{xx}$ for
all $\theta$  as it follows $R_{xx} (\theta) = A - C \cos^{2}
(\theta - \theta_{0})$, with $\theta_{0} = -8.4^{0}$. Thus, the
greatest radiation-induced $R_{xx}$ oscillatory response occurs
here when the antenna is approximately parallel or anti-parallel
to the Hall bar long-axis.\cite{grid22e} Here, it is worth noting
that the period and the phase of the radiation-induced
magnetoresistance oscillations appear not to be influenced by
$\theta$, although the amplitude of the oscillatory response is
strongly sensitive to it.

Next, we compare experimental results obtained under magnetic
field reversal. Thus, Fig. 3(a) shows the $R_{xx}$ vs. $B$ with $f = 40 GHz$ over the $B$-range $-0.25 \le B \le 0.25 T$. These data are exhibited to compare the relative extremal angular response for positive and negative $B$. As in Fig. 2, extrema of interest have been labelled in Fig. 3(a),
here as $P^{+}1$, $V^{+}1$ and $P^{+}2$ for those in the domain $B
> 0$, and $P^{-} 1$, $V^{-}1$ and $P^{-}2$ for the extrema in the
domain $B < 0$. As in Fig. 2, the angular response of the extrema
can be fit with $R_{xx} (\theta) = A \pm C \cos^{2} (\theta -
\theta_{0})$. However, the fit extracted $\theta_{0}$ here differ
substantially from zero, well beyond experimental uncertainty.
Indeed, a close inspection suggests that the $\theta_{0}$ depends
upon the magnetic field $B$ and its orientation $sgn(B)$. For
example, we find that $\theta_{0} = 64.4^{0}$ for $P^{-}1$ and
$\theta_{0} = 47^{0}$ for $P^{+}1$. Such a large difference in
$\theta_{0}$ due to magnetic field reversal is unexpected. Here,
note that since the MW antenna is far from the magnet, and well
isolated from the magnetic field, the magnetic field is not
expected to influence the polarization of the microwaves at
launch. Further, the stainless steel microwave waveguide is not
known to (and we have also not seen it) provide a microwave
frequency, magnetic field, and magnetic-field-orientation
dependent rotation to the microwave polarization. Thus, the
$\theta_{0}$ shift depending on $B$ and $sgn(B)$ looks to be a
sample effect.

Next, the role of the microwave power in the polarization
sensitivity is examined in Fig. 4. Fig. 4(a) exhibits, for $f = 37
GHz$, $R_{xx}$ vs. $B$ with $P = 0.32 mW$, along with the dark
curve. At the principal maximum $P1$ and the principal minimum
$V1$, we examine the variation of $R_{xx}$ with $\theta$, for
different values of $P$. Fig. 4(b) shows $R_{xx}$ vs. $\theta$ at
$P1$ with $P = 0.32, 1.0,$ and $3.16 mW$, and Fig. 4(c) shows the
same at $V1$. Note that $\theta_{0} = 37^{0}$ for $P1$ here at
$f=37GHz$, which differs from the $\theta_{0} = -6.7^{0}$  for
$P1$ observed at $f=39 GHz$ (see Fig. 2), and $\theta_{0} =
47^{0}$ for $P^{+}1$ at $f = 40 GHz$ (see Fig. 3). Yet, Fig. 4(b)
shows that the $\theta_{0}$ does not change with the microwave
power $P$. At $P = 0.32 mW$ in Fig. 4(c), $R_{xx}$ exhibits simple
sinusoidal variation at the $V1$ minimum, as in Fig. 2 and Fig. 3.
However, at $P=3.16 mW$, new peaks appear in Fig. 4(c) [but not in
Fig. 4(b)], in the vicinity of $\theta = 45^{0}$ and $\theta =
225^{0}$, where none were evident in the $P= 0.32 mW$ trace.

The data exhibited above showed that the phase shift, $\theta_{0}$
can vary with $f$, $B$, and sign of $B$. Next, we report results
obtained on either sides of the Hall bar device, and compare
$\theta_{0}$ obtained by measuring the angular dependence of the
$R_{xx}$. Note that these measurements were carried out on a Hall
bar device oriented perpendicular to the microwave antenna at the
outset. Thus,  the starting angle for $R_{xx}$ vs. $\theta$
measurements is $-90^{0}$ (see Fig. 5). At the top of Fig. 5, the
dark and photoexcited $R_{xx}(B)$ response of the left side of the
Hall device, $R^{L}_{xx}$ [see Fig. 5(a)] and the right side of
the Hall bar device, $R^{R}_{xx}$ [see Fig. 5(b)] are shown at
$f=43GHz$ with $P=0.5mW$ and $\theta=-90^{0}$. Here, once again,
$\theta=-90^{0}$ indicates that the MW-antenna is perpendicular to
the long axis of the Hall bar. The $R^{L} _{xx}$ vs. $\theta$
traces for $f=43GHz$ and $P=0.32mW$ at the first (P1) and second
(P2) maxima are shown in Fig. 5(c) and 5(g), respectively.
Similarly, the $R^{R} _{xx}$ vs. $\theta$ traces for $f=43GHz$ and
$P=0.32mW$ at P1 and P2 are shown in Fig. 5(d) and 5(h),
respectively. According to Fig. 5(c) and 5(d), $\theta_{0}$ for
$R^{L} _{xx}$ and $R^{R} _{xx}$ are $8.9^{0}$ and $1.6^{0}$,
respectively, at P1, and they are $-5.2^{0}$ and $-5.3^{0}$,
respectively, at P2. Also, the $R_{xx}$ vs. $\theta$ at V1 for
both sides of the sample [Fig. 5(e) and 5(f)] reveal that
$\theta_{0} = 1.7^{0}$ for $R^{L} _{xx}$ and $\theta_{0} =
-0.5^{0}$ for $R^{R} _{xx}$. Comparison of the $\theta_{0}$ values
at different extrema on either sides of the Hall bar device
indicate that the values are similar for both sides of the device
within the experimental uncertainty [see Fig. 1(e)].

\section{Discussion}

The main features in the exhibited results are therefore: (a) At
low $P$, $R_{xx} (\theta) = A \pm C \cos^{2}(\theta - \theta_{0})$
vs. the linear polarization rotation angle, $\theta$, with the
$'+'$ and $'-'$ cases describing the oscillatory maxima and
minima, respectively, see Fig. 2, 3, and 4. (b) The phase shift in
the $R_{xx} (\theta)$ response, i.e., $\theta_{0}$, varies with
$f$, $B$, and the sign of $B$ (compare Figs. 2 , 3 and 4). Yet,
$\theta_{0}$ appears insensitive to the microwave power (see Fig.
4). (c) At higher radiation power, the principal resistance
minimum exhibits additional extrema vs. $\theta$ [see Fig. 4(c)].
Point (a) demonstrates a strong sensitivity in the
radiation-induced magnetoresistance oscillations to the sense of
linear microwave polarization, in qualitative agreement with the
radiation driven electron orbit model when $\gamma< \omega = 2\pi
f$\cite{grid37, grid102}. Such sinusoidal variation of the
amplitude of the radiation induced magnetoresistance oscillations
could also be consistent with the non-parabolicity model (see Fig.
1 of Ref. \cite{grid101}). As already mentioned, the displacement
model also suggests a linear polarization
sensitivity.\cite{grid25} Consequently, the polarization angle
dependence reported here can be considered to be consistent with
the displacement model as well. Yet, the experimental feature that
the oscillations do not vanish completely at $\theta = 90^{0}$
[see, for example, Fig. 2(b), (c), and (d)] seems not to rule out
the existence of a linear-polarization-immune-term in the
radiation-induced transport. Points (b) and (c) mentioned above
are also interesting. One might also try to understand point (b),
for example, in the displacement model. Here, polarization
sensitivity\cite{grid25} is due to the inter-Landau level
contribution to the photo-current. In these experiments, the
orientation of $E_{\omega}$ is set by the antenna within the
uncertainty indicated in Fig. 1(e). The orientation of $E_{DC}$ is
variable and set by the $B$-dependent Hall angle, $\theta_{H} =
tan^{-1}(\sigma_{xy}/\sigma_{xx})$, with respect to the Hall bar
long-axis. If a particular orientation between $E_{\omega}$ and
$E_{DC}$ is preferred, say, e.g. $E_{\omega}$ $\perp$ $E_{DC}$ or
$E_{\omega}$ $//$ $E_{DC}$, for realizing large radiation-induced
magnetoresistance oscillations, and the Hall angle changes with
$B$, then a non-zero $\theta_{0}$ and a variation in $\theta_{0}$
with $B$ might be expected. However, the observed variations in
$\theta_{0}$ seem much greater than expectations since $\theta_{H}
\approx 90^{0}$ in this regime. The change in $\theta_{0}$ upon
$B$-reversal is also unexpected, and this feature identifies a
possible reason for the asymmetry in the amplitude of $R_{xx}$
under $B$ reversal often observed in such experiments. Consider
the typical $R_{xx}$ vs. $B$ measurement sweep, which occurs at a
fixed $\theta$. If peak response occurs at different $\theta_{0}$
for the two field directions, then the oscillatory $R_{xx}$
amplitudes would not be the same for positive and negative $B$.
The observed $\theta_{0}$ variations seem to suggest an effective
microwave polarization rotation in the self-response of the
photoexcited Hall bar electron device. Since $\theta_{0} \approx
\pi/4$, see Fig. 3 and 4, $B \approx 0.1 T$, and the thickness of
the 2DES lies in the range of tens of nanometers, such a scenario
would suggest giant effective polarization rotation in this high
mobility 2DES.

Finally, we reconcile our observations with other reports on this
topic.\cite{grid13, grid20} Ref. \cite{grid13} reported circular
and linear polarization immunity in the radiation-induced
magneto-resistance oscillations. Their measurements were carried
out on $4 \times 4 mm^2$ square shaped specimens, with
width-to-length ratio of one.\cite{grid13} In such a square shaped
specimen with point contacts, the current stream lines are
expected to point in different directions over the face of the
sample. Then, the variable angle between the linear microwave
polarization and the local current orientation could possibly
serve to produce an effectively polarization averaged measurement,
leading to apparent linear polarization immunity. Ref.
\cite{grid20} examined the interference of magnetointersubband
oscillations and the microwave radiation-induced
magneto-resistance oscillations, and suggested a polarization
immunity in the observed interference effect. Since the effect
examined by Wiedmann et al.\cite{grid20} differs substantially
from the conventional radiation-induced magnetoresistance
oscillations, we subscribe to the opinion that there need not be
an obvious contradiction that needs to be addressed here. At the
same time, we note that some experimental details, such as sample
geometry and the method for changing the polarization, are needed
to make a further meaningful comparison. Finally, measurements
carried out on L-shaped specimens \cite{grid5} led to the
conclusion that the phase and the period of the microwave-induced
magnetoresistance oscillations are independent of the relative
orientation of the microwave polarization and the
current\cite{grid5}, and this observation is consistent with the
initial report,\cite{grid1} and the results reported here.

\section{Conclusion}

In conclusion, experiments identify a strong sinusoidal variation
in the diagonal resistance $R_{xx}$ vs. $\theta$, the polarization
rotation angle, at the oscillatory extrema of the microwave
radiation-induced magnetoresistance
oscillations.\cite{grid23,grid25,grid101, grid37,grid102} The
results provide new evidence for the linear polarization
sensitivity in the amplitude of the radiation-induced
magnetoresistance oscillations.

\section{Acknowledgments}

Work at GSU (USA) is supported by the U.S. Department of Energy,
Office of Basic Energy Sciences, Material Sciences and Engineering
Division under DE-SC0001762, and by D. Woolard and the ARO under
W911NF-07-01-015. Work at EPS-UCIII (Spain) is supported by the
MCYT under MAT2011-24331 (Spain) and by the ITN Grant 234970 (EU).



\newpage

\end{document}